\documentstyle[prl,aps,preprint,tighten,floats,aps,epsfig,psfig,amsfonts]{revtex}

\def\btabl{\begin{table}}   \def\etabl{\end{table}}
\def\bea{\begin{eqnarray}}   \def\eea{\end{eqnarray}}
\def\bnn{\begin{eqnarray*}}   \def\enn{\end{eqnarray*}}
\def\beq{\begin{equation}}   \def\eeq{\end{equation}}  
\def\btabu{\begin{tabular}}   \def\etabu{\end{tabular}}
\def\bec{\begin{displaymath}} \def\eec{\end{displaymath}}

\def\eqref#1{(\ref{#1})}

\begin{document}
\draft
\preprint{\vbox{\baselineskip=13pt
\rightline{ENSK-ph/97-04}
\rightline{cond-mat/9802271}}}
\title{Exact four-spinon dynamical correlation function\\
in isotropic Heisenberg model}
\author{A. Abada$^{1}$, A.H. Bougourzi$^2$,  B. Si-Lakhal$^1$
 and  S. Seba$^1$ \footnote{
e-mail: enskppps@ist.cerist.dz, bougourz@alcor.concordia.ca.}} 
\vskip -0.5cm
\vspace{-0.5cm}
\address{{\small 
$^1$ D\'epartement de Physique, \'Ecole Normale Sup\'erieure \\
PB 92 Vieux-Kouba, 16050 Alger, Algeria.\\
$^2$ Department of Mathematics and Physics, Concordia University\\
7141 Sherbrooke St.~West, Montreal, Qu\'ebec H4B 1R6,
Canada.}}
\date{\today}


\maketitle \begin{abstract}
We discuss some properties of the exact four-spinon dynamical correlation
function $S_4$ in the antiferromagnetic spin 1/2 $XXX$-model, the expression
of which we derived recently. We show that the region in which it is not
identically zero is different from and larger than the spin-wave continuum.
We describe its behavior as a function of the neutron momentum transfer $k$
for fixed values of the neutron energy $\omega$ and compare it to the one
corresponding to the exact two-spinon dynamical correlation function $S_2$.
We show that the overall shapes are quite similar, even though the expression
of $S_4$ is much more involved than that of $S_2$. We finish with concluding
remarks.
\end{abstract}
\pacs{PACS: 75  78  61  67}
\newpage
\section{Introduction}

Quantum spin chains have been the subject of intensive study during the past
seven decades \cite{PSD}. Experimentally, their properties are investigated
via inelastic neutron scattering on (anti)ferromagnetic quasi
one-dimensional compounds \cite{Exp}. One important quantity that holds much
of the information related to such compounds is the dynamical correlation
function (DCF) $S$ of two local spin operators. Indeed, the neutron
scattering intensity is directly proportional to it, see for example \cite
{Exp}.

An important feature of these chains is that some of them, like the
Heisenberg model, are amenable to exact theoretical treatment while still
describing nontrivial interactions \cite{PSD}, see also \cite{Baxter}. This
is because they incorporate in them a rich mathematical structure: the
quantum affine algebra. The early work on such models consisted in
determining exactly their static thermodynamic properties, see \cite{PSD},
whereas, more recently, a fuller exploitation of the quantum symmetry using
bosonization techniques has allowed for a more systematic description of
their dynamical properties \cite{CorFun}. A systematic account of this work
is given in \cite{JimMi}.

However, one aspect of these new techniques is that the exact correlation
functions are usually obtained in the form of quite complicated contour
integrals and this renders their manipulation somewhat cumbersome. But on
the other hand, before this exact treatment, the approach to these dynamical
quantities was only approximate. Indeed, if one considers for instance the
DCF in the antiferromagnetic Heisenberg model, the focus has for a long time
been only on what we now know to be its two-spinon contribution $S_2$. First
there has been the Anderson (semi-classical) spin-wave theory \cite{And}, an
approach based on an expansion in powers of $1/s$, where $s$ is the spin of
the system and hence, is exact only in the classical limit $s=\infty $. It
can describe with some satisfaction compounds with higher spins \cite{TMMC},
but fails in the quantum limit $s=\frac 12$. Then there has been the
so-called M\"uller ansatz \cite{Muller}, which gives an approximate
expression for $S_2$ that can account for many aspects of the phenomenology
for $s=\frac 12$ compounds. Only very recently could we get for this system
a final exact expression for $S_2$ \cite{BCK}, and that gave a better
account of the data \cite{KMB}.

Now for the spin-$\frac 12$ antiferromagnetic Heisenberg model, there is a
need to go beyond the two-spinon contribution. The need is two-fold. First,
it is important to see if one is able to get useful information from these
complicated and compact expressions we alluded to above. Second and more
important perhaps is the fact that, though the exact two-spinon contribution
accounts for much of the phenomenology as we said, about 70\%, in a sense
that will become clear later in section 3, it still doesn't account for {\it %
all} of it. This point is demonstrated in particular in \cite{FMK}.

The natural step forward is to look into the exact four-spinon contribution.
To the best of our knowledge, refs \cite{Boug} and \cite{ABS} constitute the
first direct attempt in this direction. A general expression for the $n$%
-spinon contribution to the DCF in the anisotropic Heisenberg model is given
in \cite{ABS} and (a still compact one) for the isotropic limit in \cite
{Boug}. In ref \cite{ABS}, we specialize to the four-spinon case and give a
discussion of some of its properties in the isotropic and Ising limits. In
particular, we show that in the isotropic limit, the one of interest here,
the exact four-spinon contribution is safe of any potential divergences.

In this work, we further the description of the four-spinon contribution $%
S_4 $. We determine the region in which $S_4$ is not identically zero and
show that it is larger than that of $S_2$. Also, we show that the behavior
of $S_4 $ as a function of the neutron momentum transfer $k$ is similar in
its overall shape to that of the corresponding $S_2$. This result is to be
contrasted with the fact that the expression of $S_4$ is a lot more involved
than that of $S_2$, see (\ref{S4}) and (\ref{S2ex}) respectively. 

This paper is organized as follows. In the next section, we briefly discuss
the antiferromagnetic spin 1/2 Heisenberg model. We describe the spinon
Hilbert-space structure and define the dynamical correlation function. In
section 3, we succinctly review the properties of the two-spinon
contribution. We give the exact expression of $S_2$ and that of the M\"uller
ansatz, and briefly compare their main features. In section 4, we give the
exact expression of $S_4$ and discuss in detail its properties we mentioned
above. The last section comprises concluding remarks.

\section{The exact DCF in the $XXX$-model}

The antiferromagnetic $s=\frac 12$ $XXX$-Heisenberg model is defined as the
isotropic limit of the $XXZ$-anisotropic Heisenberg Hamiltonian:

\begin{equation}
H_{XXZ}=-\frac 12\sum_{n=-\infty }^\infty (\sigma _n^x\sigma _{n+1}^x+\sigma
_n^y\sigma _{n+1}^y+\Delta \sigma _n^z\sigma _{n+1}^z)\;,  \label{ham}
\end{equation}
where $\Delta =(q+q^{-1})/2$ is the anisotropy parameter. The isotropic
antiferromagnetic limit is obtained via the limit $q\rightarrow -1^{-}$ or
equivalently $\Delta \rightarrow -1^{-}$. Here $\sigma _n^{x,y,z}$ are the
usual Pauli matrices acting at the site $n.$ The exact diagonalization
directly in the thermodynamic limit of the Hamiltonian in (\ref{ham}) is
possible using the $U_q(\widehat{sl(2)})$ quantum group symmetry present in
the model \cite{JimMi}. The resulting Hilbert space ${\cal F}$ consists of $%
n $-spinon energy eigenstates $|\xi _1,...,\xi _n\rangle _{\epsilon
_1,...,\epsilon _n;\,i}$ built on the two vacuum states $|0\rangle _i$, $%
i=0,1$ such that: 
\begin{equation}
H_{XXZ}|\xi _1,...,\xi _n\rangle _{\epsilon _1,...,\epsilon
_n;\,i}=\sum_{j=1}^ne(\xi j)|\xi _1,...,\xi _n\rangle _{\epsilon
_1,...,\epsilon _n;\,i\;},
\end{equation}
where $e(\xi j)$ is the energy of spinon $j$ and $\xi _j$ is a spectral
parameter living on the unit circle. In the above equation, $\epsilon _j=\pm
1$ and the index $i$ refers to the boundary condition on the spin chain, see 
\cite{JimMi}. The translation operator $T$ which shifts the spin chain by
one site acts on the spinon eigenstates in the following manner: 
\begin{equation}
T|\xi _1,...,\xi _n\rangle _{\epsilon _1,...,\epsilon
_n;\,i}=\prod_{i=1}^n\tau (\xi _i)|\xi _1,...,\xi _n\rangle _{\epsilon
_1,...,\epsilon _n;1-i\;},
\end{equation}
where $\tau (\xi _j)=e^{-ip_j}$ and $p_j$ is the lattice momentum of spinon $%
j$. The expressions of the spinon energy and lattice momentum in terms of
the spectral parameter are quite cumbersome in the anisotropic case, see 
\cite{JimMi,ABS}, but simplify considerably in the isotropic limit, see eq (%
\ref{DisRel}) below. The completeness relation in ${\cal F}$ reads: 
\begin{equation}
{\bf I}=\sum_{i=0,1}\sum_{n\geq 0}\sum_{\{\epsilon _j=\pm 1\}_{j=1,n}}\frac 1%
{n!}\oint \prod_{j=1}^n\frac{d\xi _j}{2\pi i\xi _j}\;|\xi _1,...,\xi
_n\rangle _{\epsilon _1,...,\epsilon _n;\,i\;i;\,\epsilon _1,...,\epsilon
_n}\langle \xi _1,...,\xi _n|\;.
\end{equation}

The two-point DCF is the Fourier transform of the zero-temperature
vacuum-to-vacuum two-point function, i.e., it is defined by: 
\begin{equation}
S^{i,+-}(\omega ,k)=\int_{-\infty }^\infty dt\sum_{m\in {\Bbb {Z}}%
}e^{i(\omega t+km)}\,_i\langle 0|\sigma _m^{+}(t)\,\sigma _0^{-}(0)|0\rangle
_i\;,
\end{equation}
where $\omega $ and $k$ are the neutron energy and momentum transfer
respectively and $\sigma ^{\pm }$ denotes $(\sigma ^x\pm i\sigma ^y)/2$. The
DCF is such that: 
\begin{equation}
S^{i,+-}(\omega ,k)=S^{i,+-}(\omega ,-k)=S^{i,+-}(\omega ,k+2\pi )\,,
\label{symDCF}
\end{equation}
the two relations that express the reflection and periodicity symmetries on
the linear chain. Inserting the completeness relation and using the
Heisenberg relation: 
\begin{equation}
\sigma _m^{x,y,z}(t)=\exp (iH_{XXZ}\,t)\,T^{-m}\sigma
_0^{x,y,z}(0)\,T^m\,\exp (-iH_{XXZ}\,t)\;,
\end{equation}
we can write the transverse DCF as the sum of $n$-spinon contributions: 
\begin{equation}
S^{i,+-}(\omega ,k)=\sum_{n\,\,{\rm even}}S_n^{i,+-}(\omega ,k)\;,
\end{equation}
where the $n$-spinon DCF $S_n$ is given by: 
\begin{eqnarray}
S_n^{i,+-}(\omega ,k) &=&\frac{2\pi }{n!}\sum_{m\in {\Bbb {Z}}%
}\,\sum_{\epsilon _1,...,\epsilon _n}\oint \prod_{j=1}^n\frac{d\xi _j}{2\pi
i\xi _j}\,e^{im(k+\sum_{j=1}^np_j)}\,\delta \left( \omega
-\sum_{j=1}^ne_j\right) \,  \nonumber \\
&&\times \;X_{\epsilon _n,...,\epsilon _1}^{i+m}(\xi _n,...,\xi
_1)\;X_{\epsilon _1,...,\epsilon _n}^{1-i}(-q\xi _1,...,-q\xi _n)\;,
\end{eqnarray}
relation in which $X^i$ denotes the form factor: 
\begin{equation}
X_{\epsilon _1,...,\epsilon _n}^i(\xi _1,...,\xi _n)\equiv \,_i\langle
0|\sigma _0^{+}\left( 0\right) |\xi _1,...,\xi _n\rangle _{\epsilon
_1,...,\epsilon _n;\,i}\;.
\end{equation}
Note that each $S_n$ must satisfy relations (\ref{symDCF}).

The exact expression of this form factor has been determined in \cite{JimMi}%
. Using this form factor, we can give an exact expression for the $n$-spinon
DCF in the anisotropic case, see \cite{ABS}, and determine exactly its
isotropic limit, see \cite{Boug}. This limit is obtained via the replacement 
\cite{JimMi,ABS}: 
\begin{equation}
\xi =ie^{-2i\varepsilon \rho }\,;\qquad q=-e^{-\varepsilon }\,,\qquad
\varepsilon \rightarrow 0^{+}\,,  \label{IsoLim}
\end{equation}
where $\rho $ is the spectral parameter suited for this limit. The
expressions of the energy $e$ and momentum $p$ in terms of $\rho $ then
read: 
\begin{equation}
e(\rho )=\frac \pi {\cosh (2\pi \rho )}=-\pi \sin \,p\;;\quad \cot \,p=\sinh
(2\pi \rho )\;;\quad -\pi \leq p\leq 0\;.  \label{DisRel}
\end{equation}
The transverse two-spinon DCF $S_2$ does not involve a contour integration
and has been given in \cite{BCK}. The four-spinon one $S_4$ involves only
one contour integration and its expression is given in \cite{ABS}. In the
next section, we review the properties of $S_2$ and in the one that follows
it, we discuss those of $S_4$.

\section{The exact two-spinon DCF}

The exact expression of the two-spinon DCF\ of the spin $\frac 12$ $XXX$%
-model is given in \cite{BCK} and reads: 
\begin{equation}
S_2^{+-}(\omega ,k-\pi )=\frac 14\frac{e^{-I(\rho )}}{\sqrt{\omega
_{2u}^2-\omega ^2}}\,\Theta (\omega -\omega _{2l})\,\Theta (\omega
_{2u}-\omega )\;,  \label{S2ex}
\end{equation}
where $\Theta $ is the Heaviside step function and the function $I(\rho )$
is given by: 
\begin{equation}
I(\rho )=\int_0^{+\infty }\frac{dt}t\frac{\cosh (2t)\,\cos (4\rho t)-1}{%
\,\sinh (2t)\cosh (t)}\,e^t\;.  \label{IdeT}
\end{equation}
$\omega _{2u(l)}$ is the upper (lower)\ bound of the two-spinon excitation
energies called the des Cloizeaux and Pearson \cite{Muller,BCK} upper
(lower) bound or limit. They read: 
\begin{equation}
\omega _{2u}=2\pi \sin \frac k2;\qquad \quad \omega _{2l}=\pi \,|\sin \,k|\;.
\label{dCP}
\end{equation}
The quantity $\rho =\rho _1-\rho _2$ and is related to $\omega $ and $k$ by
the relation: 
\begin{equation}
\cosh \,\pi \rho =\sqrt{\frac{\omega _{2u}^2-\omega _{2l}^2}{\omega
^2-\omega _{2l}^2}}\;,
\end{equation}
a relation obtained using eq (\ref{DisRel}) and the energy-momentum
conservation laws: 
\begin{equation}
\omega =e_1+e_2\,;\qquad \quad k=-p_1-p_2\;.
\end{equation}
Note that the explicit expression of $S_2$ given in eq (\ref{S2ex})
satisfies the reflection and periodicity symmetries expressed in (\ref
{symDCF}).

The properties of $S_2$ have been discussed in \cite{KMB}. There, a
comparison with the M\"uller ansatz \cite{Muller} is given. This latter was
derived from the properties of some solutions to the Bethe-ansatz equations,
from numerical calculations on finite spin chains and from an analysis of
phenomenological results. It reads: 
\begin{equation}
S_2^{(a)}(\omega ,k-\pi )=\frac A{2\pi }\frac{\Theta (\omega -\omega
_{2l})\,\Theta (\omega _{2u}-\omega )}{\sqrt{\omega ^2-\omega _{2l}^{\,2}}}%
\;,  \label{S2Mul}
\end{equation}
where $A$ is a constant determined in such a way to fit best the
phenomenology \cite{Muller}. There are two main differences between the
exact expression (\ref{S2ex}) and the approximate one (\ref{S2Mul}), \cite
{KMB}. First, the two-spinon threshold at $\omega _{2l}$ is more singular in
(\ref{S2ex})\ than in (\ref{S2Mul}). Second, at the upper two-spinon
boundary $\omega _{2u}$, $S_2$ vanishes smoothly whereas $S_2^{(a)}$ has a
sharp cut-off. But if one defines the frequency moments of the DCF: 
\begin{equation}
K_n(k)=\int_{-\infty }^{+\infty }d\omega \,\omega ^n\,S(\omega ,k)\;,
\label{fremom}
\end{equation}
one shows that as $k\rightarrow 0,$ the moment of $S_2$ vanishes as: 
\begin{equation}
K_n^{(2)}(k)\sim \omega _{2u}^{n+1}(k)\;,
\end{equation}
and the same holds for the M\"uller ansatz $S_2^{(a)}$.

Actually, the frequency moments (\ref{fremom}) are particular cases of a set
of sum rules the DCF is known to satisfy exactly. For
exemple, we know that the first moment is exactly equal to \cite{HohBri}: 
\begin{equation}
K_1(k)=\frac{4\ln \,2-1}6\,(1-\cos \,k)\;.
\end{equation}
It turns out that the same frequency moment for $S_2$ is such that \cite{KMB}%
: 
\begin{equation}
\frac{K_1^{(2)}(k)}{K_1(k)}\simeq 70\%\;.  \label{80per}
\end{equation}
This means that, according to this sum rule, $S_2$ is way off the total DCF $%
S$ by roughly 30\%. In fact, other exact sum rules confirm this trend, see 
\cite{KMB}. Of course, those remaining 30\% are ``filled'', so to speak, by
the $n>2$-spinon contributions. The natural question that comes to mind is:
how much $S_4$ takes up from these 30\%? Obviously, one has first to study
the behavior of this contribution before trying to answer this question,
and this we do in the remainder of this paper.

\section{ The exact four-spinon DCF}

The analytic expression of the four-spinon DCF has been worked out in \cite
{ABS} and for $0\leq k\leq \pi $ it reads: 
\begin{equation}
S_4^{+-}(\omega ,k-\pi )=C_4\int_{-\pi }^0dp_3\int_{-\pi }^0dp_4\,F(\rho
_1,...,\rho _4)\;,  \label{S4}
\end{equation}
where $C_4$ is a numerical constant and the integrand $F$ is given by: 
\begin{equation}
F(\rho _1,...,\rho _4)=\sum_{(p_1,\,p_2)}\frac{f(\rho _1,...,\rho
_4)\,\sum_{\ell =1}^4|g_\ell (\rho _1,...,\rho _4)|^2}{\sqrt{W_u^2-W^2}}\;.
\label{FS4}
\end{equation}
The different quantities involved in this expression are: 
\begin{eqnarray}
W &=&\omega +\pi \,(\sin \,p_3+\sin \,p_4\,)\;;  \nonumber \\
W_u &=&2\pi \left| \sin \,\frac K2\right| \;;  \nonumber \\
K &=&k+p_3+p_4\;;  \nonumber \\
\cot \,p_j &=&\sinh (2\pi \rho _j)\;,\qquad -\pi \leq p_j\leq 0\;.
\label{not}
\end{eqnarray}
The function $f$ is given by: 
\begin{equation}
f(\rho _1,...,\rho _4)=\exp \,\left[ -\sum_{1\leq j<j^{^{\prime }}\leq
4}I\,(\rho_{j j'})\right] \;,  \label{f}
\end{equation}
where $\rho_{j j'}=\rho _j-\rho _{j'}$ and the function $g_\ell $ reads: 
\begin{eqnarray}
g_\ell &=&(-1)^{\ell +1}(2\pi )^4\sum_{j=1}^4\cosh \,\,(2\pi \rho _j) 
\nonumber \\
&&\times \sum_{m=\Theta (j-\ell )}^\infty \frac{\prod_{i\neq \ell }(m-\frac 1%
2\Theta (\ell -i)+i\rho _{ji})}{\prod_{i\neq j}\pi ^{-1}\sinh (\pi \rho
_{ji})}\prod_{i=1}^4\frac{\Gamma (m-\frac 12+i\rho _{ji})}{\Gamma (m+1+i\rho
_{ji})}\;,  \label{gl}
\end{eqnarray}
where $\Theta $ is the Heaviside step function. In (\ref{FS4}),
the sum $\sum_{(p_1,p_2)}$ is over the two pairs $%
(p_1,p_2)$ and $(p_2,p_1)$ solutions of the energy-momentum conservation
laws: 
\begin{equation}
W=-\pi (\sin \,p_1+\sin \,p_2)\;;\qquad \quad K=-p_1-p_2\;.  \label{CL}
\end{equation}
They read: 
\begin{equation}
(p_1,p_2)=\left( -\frac K2+\arccos \left( \frac W{2\pi \sin \frac K2}\right)
\;,\quad -\frac K2-\arccos \left( \frac W{2\pi \sin \frac K2}\right) \right)
\;.  \label{SCL}
\end{equation}
Note that the solution in (\ref{SCL}) is allowed as long as $W_l\leq W\leq
W_u$ where $W_u$ is given in (\ref{not}) and: 
\begin{equation}
W_l=\pi |\sin \,K|\;.
\end{equation}

We henceforth put ourselves in the interval $0\leq k\leq \pi$. To get
the behavior of $S_4$ for the values of $k$ outside this interval,
one uses the symmetry relations (\ref{symDCF}) given in section 2.
In the work \cite{ABS}, we have discussed the behavior of the function $F$
given in (\ref{FS4}). We have shown that the series $g_\ell $ is convergent.
We have also shown that in the region where two $\rho _i$'s or more get
equal, the function $g_\ell $ is finite. The function $f$ going to zero in
these same regions \cite{KMB}, this means the integrand $F$ of $S_4$ has a
nice regular behavior there. Furthermore, we have shown that $F$ is
exponentially convergent when one of the $\rho _i$'s goes to infinity, which
means the two integrals over $p_3$ and $p_4$ in (\ref{S4})\ do not yield
infinities. All these analytic results pave the way to ``safe'' numerical
manipulations.

The first thing we wish to discuss in this work is the extent of
the region in the $(k,\omega )$-plane in which $S_4$ is not
identically zero, and compare it to that of $S_2$. Remember that from
(\ref{S2ex}), $S_2$ is zero identically outside the spin-wave continuum
$\omega _{2l}(k)\leq \omega \leq \omega _{2u}(k)$, where $\omega _{2l,u}(k)$
are given in (\ref{dCP}). From the condition $W_l\leq W\leq W_u$ discussed
after eq (\ref{SCL}), one infers that in order for $S_4$ to be nonzero
identically, one has to have $\omega _{4l}\leq \omega \leq \omega _{4u}$,
where: 
\begin{eqnarray}
\omega _{4l}(k) &=&3\pi \sin (k/3)\quad {\rm for}\quad 0\leq k\leq \pi /2\;;
\nonumber \\
\omega _{4l}(k) &=&3\pi \sin (k/3+2\pi /3)\quad {\rm for}\quad \pi /2\leq
k\leq \pi \;;  \nonumber \\
\omega _{4u}(k) &=&4\pi \cos (k/4)\quad {\rm for}\quad 0\leq k\leq \pi \;.
\label{dCP4}
\end{eqnarray}
All these branches are plotted in fig.~1.

The first thing we immediately see is that the $S_4$-region, i.e., the
region in which $S_4$ is not identically zero, is not confined to the
spin-wave continuum delimited by the dCP branches $\omega _{2l,u}$ given in (%
\ref{dCP}). This means that, a fortiori, the full $S$ is also not confined
to the $S_2$-region. This fact is confirmed by early finite chain numerical
calculations \cite{Muller} and the phenomenology \cite{Exp}.
Furthermore, fig.~1 shows that for $0\leq k/\pi \leq 1/2$, the $S_4$-region
is entirely beyond the $S_2$-region. This means that for this interval, we
may expect $S_2$ to be dominant in $S$ within the spin-wave continuum.
However, for $1/2\leq k/\pi \leq 1$, there is overlap between the two
regions such that the $S_2$-region is more or less within the $S_4$-region.
We may therefore expect here the contribution of $S_4$ to play a r\^ole, and
hence, we expect $S_2$ to be a little less dominant within the spin-wave
continuum.

The next feature we discuss in this work is the behavior of $S_4$ as a
function of $k$ for fixed $\omega $. Figs.~2a, 3a and 4a show the behavior
of $S_4$ for $\omega /\pi =0.45,0.5$ and $0.75$ respectively. Note that we
have scaled $S_4$ to appropriate units. Figs.~2b, 3b and 4b show the
behavior of $S_2$ as a function of $k$ for the same values of $\omega $.
Let us for example discuss the case $\omega /\pi =1/2\,$. We see from fig.~3b
that the function $S_2$ vanishes for (roughly)\ $k/\pi \leq 0.8$. Looking
back into fig.~1, this corresponds indeed to the region outside the spin-wave
continum for $\omega /\pi =1/2$, i.e., $k/\pi \leq 5/6$. The function $S_2$
starts at $k/\pi =5/6$ with a large value and goes to a minimum at $k/\pi =$
1.

Fig.~3a shows that the function $S_4$ has a somewhat similar behavior. From
the figure, we read that $S_4$ too is not vanishing for (roughly) $0.8\leq
k/\pi \leq 1$\thinspace . From fig.~1, that corresponds within the $S_4$%
-region to $0.84\leq k/\pi \leq 1$. But fig.~1 shows also that for $\omega
/\pi =0.5$, $S_4$ may be non-vanishing in the interval $0\leq k/\pi \leq 0.16
$. That contribution doesn't appear on fig.~3a, presumably because $S_4$
there is negligeable. This is confirmed in the case $\omega /\pi =0.45$
which is close to the case $\omega /\pi =0.5$: there we see $S_4$ having a
very small contribution in the corresponding interval. From fig.~3a, we see
that $S_4$, like $S_2$, starts from a large value at its lower boundary
$k/\pi =0.84$ and decreases while moving to $k/\pi =1$.

The case $\omega /\pi =0.75$ is practically the same.
Fig.~4b shows $S_2$ starting from
zero at roughly $k/\pi =0.25$, getting to a maximum and sharply dropping to
zero a little further. It then starts sharply again from a large value a
little after $k/\pi =0.7$ and decreases to a minimum at $k/\pi =1$. This is
also consistent with fig.~1: the function $S_2$ starts to be nonvanishing
for $\omega /\pi =0.75$ at $k/\pi =0.2447$. It stays nonvanishing until $%
k/\pi $ reaches the value $0.2699.$ It remains identically zero until $k/\pi 
$ reaches the value $0.7301$ at which we enter back into the $S_2$-region.

As we said, $S_4$ has in this case too the same overall
behavior as that of $S_2$. In fig.~4a, we see that $S_4$ starts to increase
from the value zero at $k=0$. It goes quickly to a maximum before dropping
sharply to zero a little before $k/\pi =0.3$. It stays at zero till a little
after $k/\pi =0.7$ and increases sharply. Then it decreases while wiggling
to $k/\pi =1$. This overall behavior is also consistent with fig.~1. Indeed,
for $\omega /\pi =0.75$, the $S_4$-region starts at $k=0$ and extends first
to $k/\pi =0.2413$. Then we get outside this region from this value of $k$
till $k/\pi =0.7587$. Then $S_4$ is no more identically zero beyond this
point until we reach the point $k=\pi$. As we said, this is quite
consistent with fig.~4a.

\section{Discussion and conclusion}

In this work, we have discussed the behavior of the exact four-spinon
contribution $S_4$ to the dynamical correlation function $S$ of the $s=1/2$
antiferromagnetic Heisenberg model and compared it to the one of the exact
two-spinon contribution $S_2$. We first reviewed the model and the spinon
structure of the corresponding Hilbert space. We then gave a brief account
of the results concerning $S_2$ and its comparison to the M\"uller ansatz.
The first thing concerning $S_4$ we discussed is the region in the
$(k,\omega )$-plane in which $S_4$ is not identically zero. We found it to be
different and larger than that of $S_2$, i.e., the spin-wave continuum. Both
regions are drawn in fig.~1. We then discussed the behavior of $S_4$ as a
function of $k$ for fixed values of $\omega $, and compared it to the one of 
$S_2$. These behaviors are plotted in figs.~2,3 and 4. We have found that
the overall shape of $S_4$ is more or less the same as that of $S_2$.

The first thing to emphasize we think is the overall similarity between the
shapes of $S_4$ and $S_2$. This is not at all expected from the outset, given
the more complicated expression of $S_4$, see (\ref{S4}), as compared to
that of $S_2$, see (\ref{S2ex}). Does this mean that the shape of
the other $n>2$-spinon contributions and hence of the total $S$ is already
more or less ``traced'' by that of $S_2$? We think that at this stage,
it is too early for such an inference: this is a preliminary
investigation into the behavior of $S_4$ and clearly more work is needed.

In any case, it would have been interesting to measure $S_4$ for other
(larger) values of $\omega$, especially in regions where $S_2$ is identically
zero whereas $S_4$ is not, see fig.~1. But as $\omega$ increases,
the structure of the function $F(\rho _1,...,\rho _4)$ of eq(\ref{FS4})
in the $(p_3,p_4)$-plane
gets ``richer'', which means numerically harder to handle. To illustrate
this point, we have plotted for the reader in figs.~5 the function $F$ for
$k/\pi =1/2$ and $\omega=2\pi$ (5a), $\omega=3\pi$ (5b). One can see that
as $\omega$ increases, the function $F$ gets distributed nontrivially in
larger areas in the $(p_3,p_4)$-plane, with an increasing more involved
structure. This is the main reason why we preferred to defer the discussion
of these regions to future work. Also, for the same reason, we have deferred
the systematic discussion of $S_4$ as a function of $\omega $ for fixed
values $k$.

One other interesting question we haven't touched on in this work but merely
alluded to at the end of section 3 is the following: how much $S_4$ accounts
for in the total $S$? In other words, is $S_2+S_4$ better an approximation
to the total $S$ than $S_2$ alone, and if yes, by how much? As we said, we
know that $S_2$ accounts for about 70\% of the total $S$, which means that
roughly 30\% are left for the $n>2$-spinon contributions. To tackle this
question as it should, one has to rely on a certain number of sum rules $S$
is known to satisfy exactly. Then one compares the contribution of $S_2+S_4$
to the exact result and carries a discussion thereon.

The other interesting question one may ask is the physical interpretation
and implications of the behavior of the four-spinon DCF we have described in
this work. What would certainly be interesting is to be able to
systematically measure $S$ outside the spin-wave continuum so that we are
assured of having the effects of the two-spinon contribution eliminated.

\begin{acknowledgments}
We would like to thank Asmaa Abada and Joaquim Matias for all their help.
This is a revised version of the work and we thank the referee for his
remark regarding the symmetry about the axis $k=\pi$.
\end{acknowledgments}

\newpage\noindent {\bf Figure captions}

\begin{description}
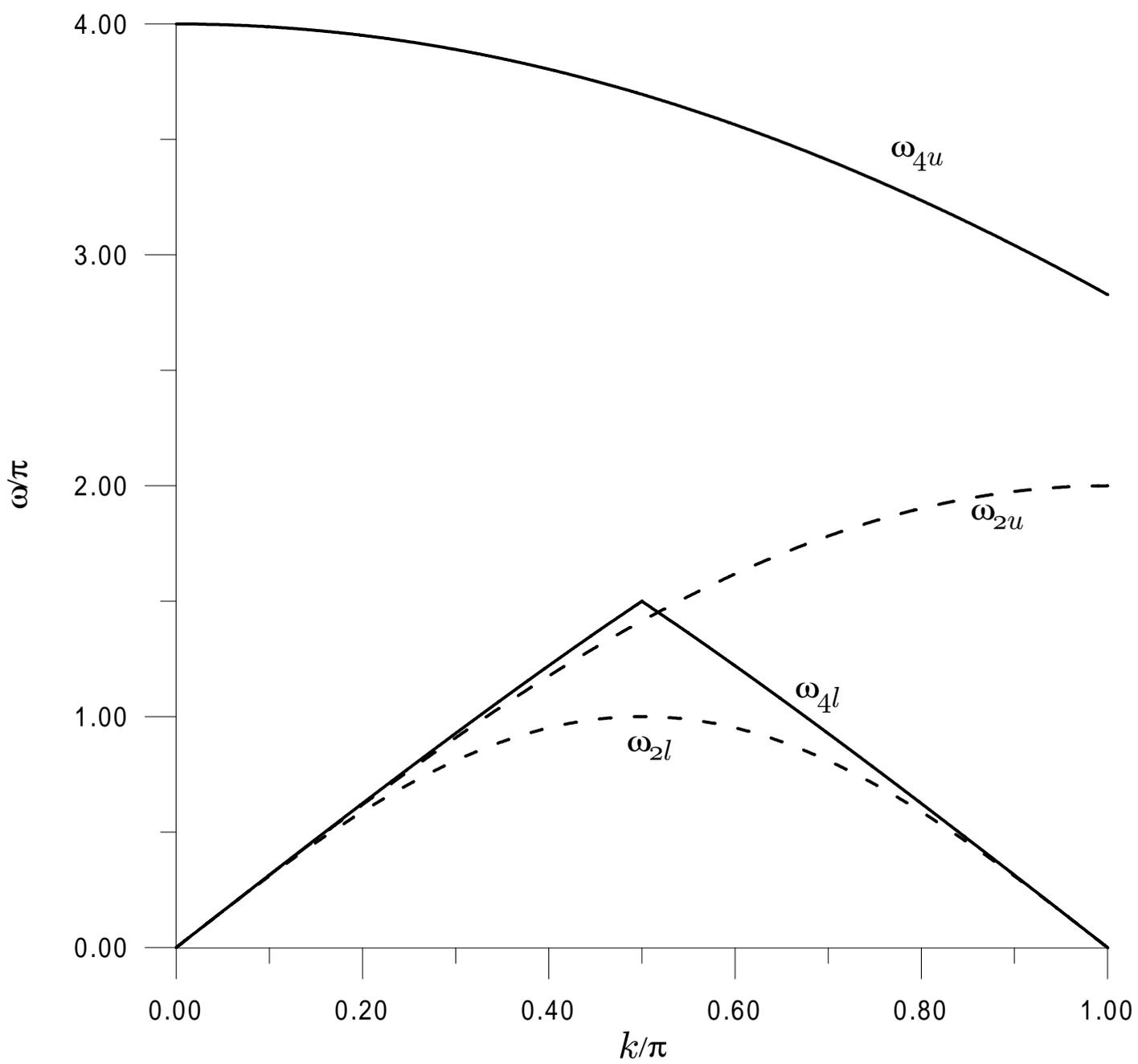
\item  Fig.~1:\ The regions in the $(k,\omega )$-plane inside which $S_2$
(dashed lines) and $S_4$ (solid lines)\ are not identically zero. The
boundaries are indicated as defined in the text.

\item  Figs.~2--4: $S_4$ (a) and $S_2$ (b) as functions of $k$ for fixed $%
\omega $. The values of $\omega $ are indicated. Note that for the sake of
illustration, $S_4$ is plotted to appropriate units.

\item  Figs.~5: The function $F(\rho _1,\rho _2,\rho _3,\rho _4)$ of
relation (\ref{S4}) plotted in the $(p_3,p_4)$-plane for $k=0.5\pi $ and $%
\omega =2\pi $ (a), $\omega =3\pi $ (b). Note that each $F$ is scaled to
appropriate units.
\end{description}

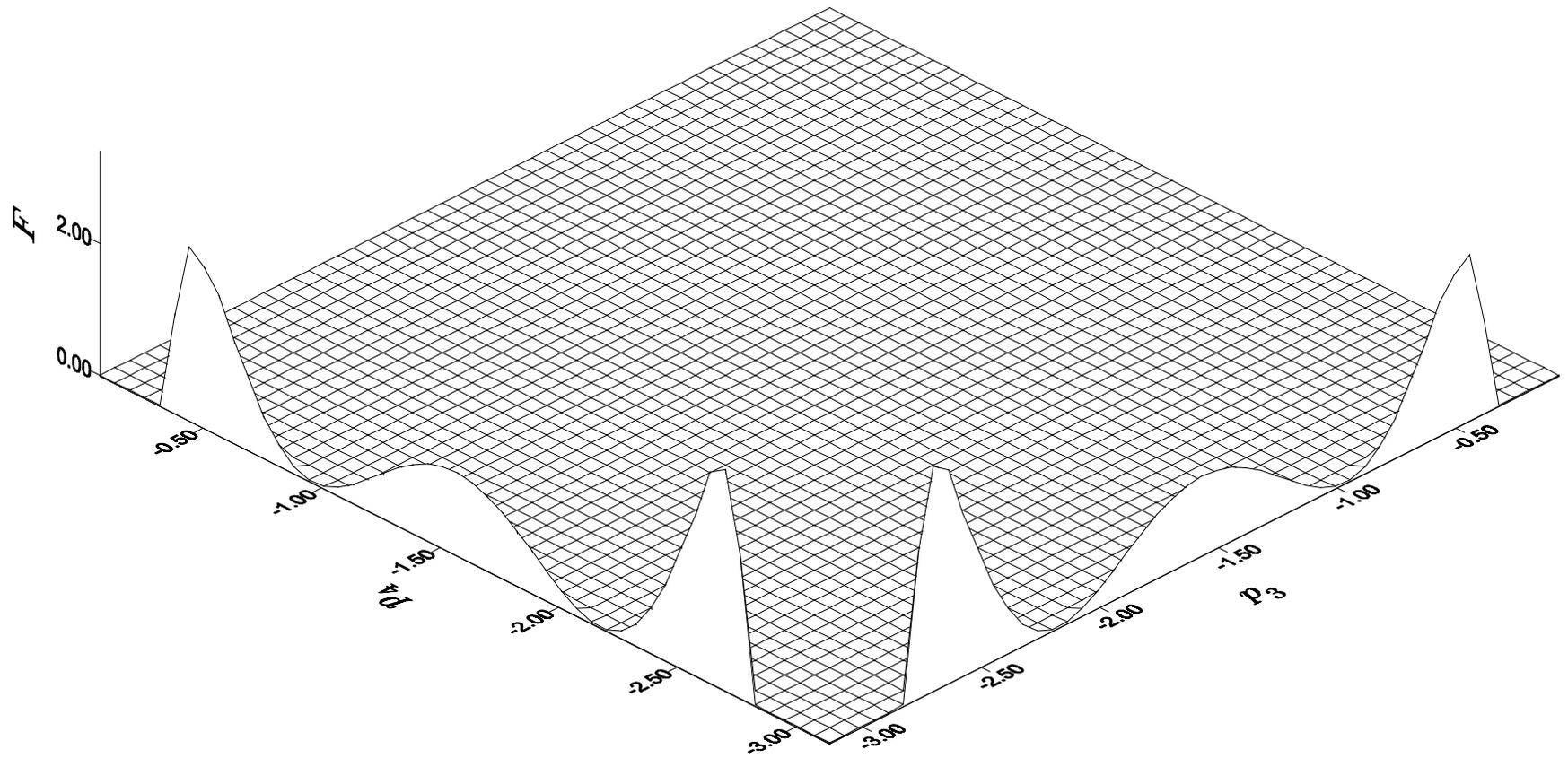
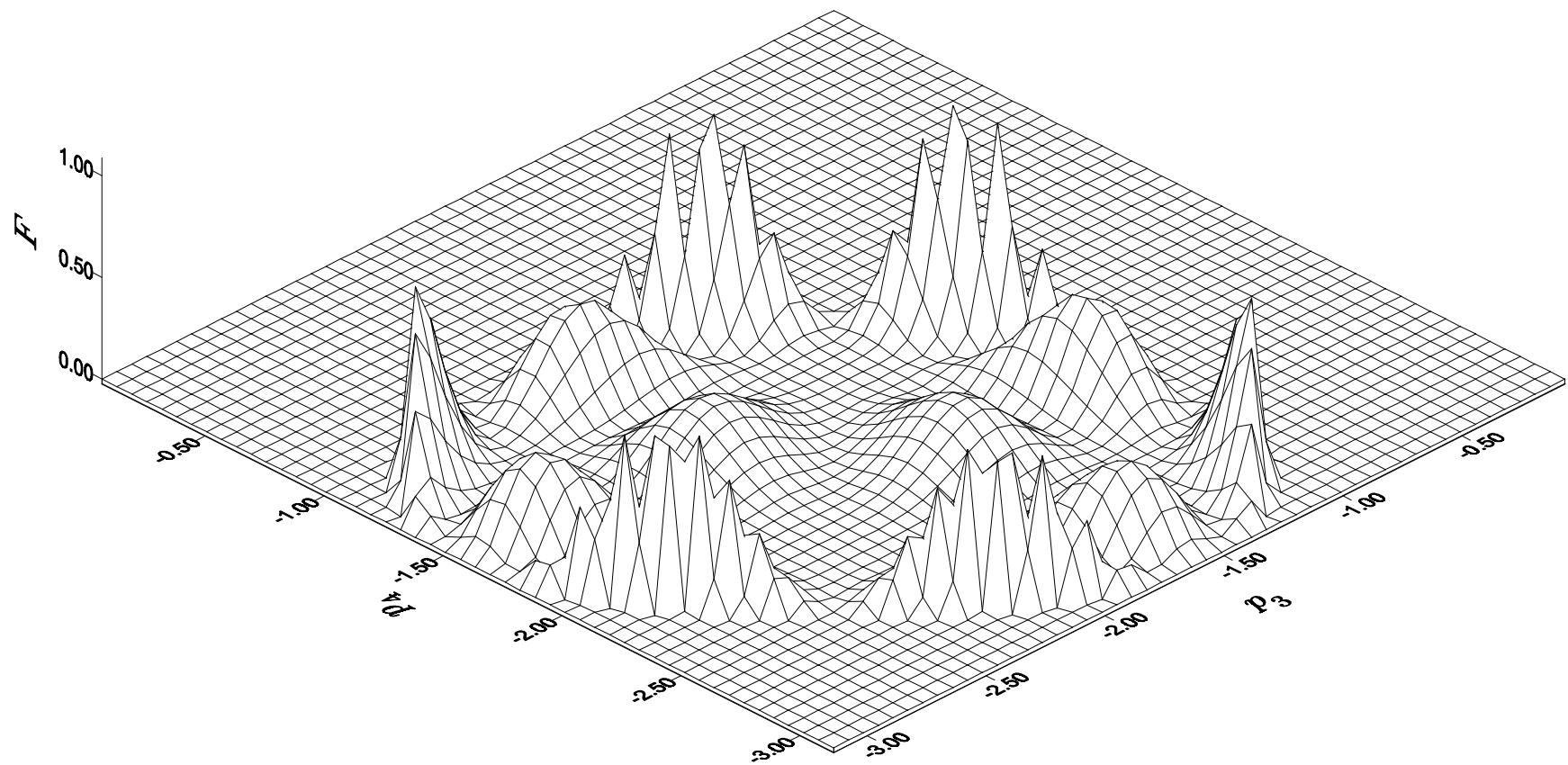
\end{document}